\documentclass[twocolumn,floats,superscriptaddress,floatfix,prb]{revtex4}
\usepackage{graphicx}
\usepackage{bm}
\usepackage{amsfonts,amssymb,amsmath}
\usepackage{graphics,psfrag,epsfig}

\begin{document}

\title{``Locally homogeneous turbulence'' Is it an inconsistent
framework?}
\author{Uriel Frisch} \author{J\'er\'emie Bec}
\affiliation{CNRS, Laboratoire Cassiop\'ee,
 Observatoire de la C\^{o}te d'Azur, BP4229, 06304 Nice Cedex 4,
 France }
\author{Erik Aurell}
\affiliation{Department of Physics, KTH, AlbaNova University Center, SE-106 91,
  Stockholm, Sweden}
\date{\today}

\begin{abstract}
In his first 1941 paper Kolmogorov assumed that the velocity has increments
which are  homogeneous and independent of the velocity
at a suitable reference point. This assumption of local homogeneity
is consistent with the nonlinear dynamics only in an asymptotic
sense when the reference point is far away. The inconsistency is illustrated
numerically using the Burgers equation. Kolmogorov's derivation of 
the four-fifths law for the third-order structure function and its anisotropic
generalization are actually valid only for homogeneous turbulence,
but a local version due to Duchon and Robert still holds. A Kolomogorov--Landau
approach is proposed to handle the effect of 
fluctuations in the large-scale velocity on small-scale statistical
properties; it is is only a mild extension of the 1941 theory and does 
not incorporate intermittency effects.
\end{abstract}

\maketitle
\vspace{2truemm}

The concept of homogeneity, which goes back to Kelvin,\cite{kelvin} applies
to flows whose statistical properties are invariant under (space) translations.
For translation-invariant equations, such as the Burgers or
Navier--Stokes equations  and in the absence of boundaries it is obvious
that homogeneity is dynamically preserved. More precisely, with a homogeneous
initial velocity field $u_0$, as long as the solution is unique, it will
also be homogeneous. 

A less restrictive condition is that of \emph{incremental
homogeneity}: we now demand only that the statistical properties of
the increments $u(x+r)- u(x)$ be invariant under translations, that is
depend only on $r$ but not on $x$. In the time domain for which
homogeneity is called stationarity, a well-known example of an
incrementally stationary random function which is not stationary is
the Brownian motion curve. As pointed out by Shiryaev,\cite{shiryaev}
Kolmogorov published in 1940, one year before his famous publications
on turbulence, two probability papers. One discusses among other
things incremental homogeneity \cite{k40a} and the other one ``Wiener
spirals'', self-similar Gaussian random functions now called
fractional Brownian motions.\cite{k40b} He was dealing with random
functions whose energy spectrum $E(k)$ has an infrared divergence,
that is $\int_0^\infty E(k)dk$ diverges at $k=0$. For example the
Brownian motion has $E(k) \propto k^{-2}$. Turbulence in fluids and
plasmas is replete with infrared divergent power-law spectra, for
example, the Kolmogorov $k^{-5/3}$ spectrum. In such cases the
variance of the velocity increment is finite but the variance of the
velocity itself is infinite.  Of course, upon introduction of an
infrared cutoff $k_0>0$, homogeneity is recovered. For example the
homogeneous Ornstein--Uhlenbeck Gaussian process which has the
spectrum $E(k) = C/(k_0^2+k^2)$ goes over into Brownian motion as
$k_0\to 0$.

In his first 1941 paper \cite{k41a} (K41) Kolmogorov pointed out the
need to work with the increments of the velocity. However, rather than
assuming that the turbulence is incrementally homogeneous, he assumed
\emph{local homogeneity}. Specifically he took a space-time reference
point $P^{\rm (0)}= ({\bm r}^{\rm (0)},\, t^{\rm (0)})$ and, for an
arbitrary space-time point $P$, considered the space-time increments
${\bm w}(P) \equiv {\bm u}(P) - {\bm u}_0$ where ${\bm u}_0 = {\bm
u}(P^{\rm (0)})$ and then defined local homogeneity as the property
that for any choice of $n$ points $P_1$, $P_2$, \ldots, $P_n$, the
joint distribution of the $n$ increments should be independent of both
$P^{\rm (0)}$ and of ${\bm u}_0$.\footnote{We have here very slightly
changed the notation since Kolmogorov was not using vector notation;
we have also corrected an obvious typographical error: Kolmogorov used
$t$ instead of $t^{\rm (0)}$ in the definition of $P^{\rm (0)}$.}  In
other words, Kolmogorov was assuming not only incremental homogeneity
and incremental stationarity, he was also assuming independence of the
velocity at the reference point.

As stressed by Hill,\cite{hill97,hill02arxiv} various definitions of
local homogeneity have been used in the literature.  For example Monin
and Yaglom \cite{my2} mostly take it to be equivalent to incremental
homogeneity but, of course, also quote Kolmogorov's full
definition. It is not clear why Kolmogorov felt it necessary to
introduce the additional assumption of independence of the velocity at
the reference point. It is conceivable that when he wrote his first
1941 turbulence paper he already knew that he would need this
assumption to avoid extra terms appearing otherwise in the derivation
of the four-fifths law.\cite{k41c} In this paper we shall discuss
successively the problems raised by incremental homogeneity and those
raised by the assumption of independence of ${\bm u}_0$.

First we observe that it is far from obvious that incremental
homogeneity is consistent with the quadratic nonlinear dynamics of the
Burgers and Navier--Stokes equation. Indeed the square of an
incrementally homogeneous function has no reason to share this
property because the increments of the square are not the squares of
the increments.

\begin{figure}[htbp]
\centerline{\psfig{file=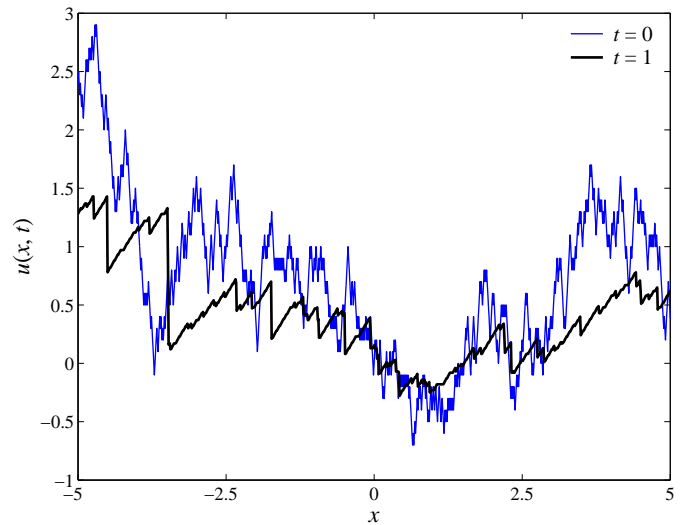,width=0.5\textwidth}}
\caption{A snapshot of a Brownian initial velocity (thin line)
and the corresponding solution to  the Burgers equation at time one 
(thick line);  notice the proliferation of shocks.}
\label{f:brown}
\end{figure}

It is easy to show the dynamical inconsistency of incremental
homogeneity using the one-dimensional Burgers equation
\begin{equation}
\partial_t u + \partial_x{u^2\over 2} = \nu\partial_{xx}u.
\label{burgers}
\end{equation}
We shall work with the solution in the limit of vanishing viscosity
which has a simple representation: writing $u = -\partial_x \psi$, we
have
\begin{equation}
\psi (x,t) = \max_a \left[\psi(a,0) -{(x-a)^2\over 2t}\right],
\label{max}
\end{equation}
where $\psi(x,0)$ is the initial (Lagrangian) velocity potential (see,
e.g.\ Ref.~\onlinecite{burgulence}). Eq.~(\ref{max}) can be
implemented numerically in a very efficient way using the Fast
Legendre transform algorithm.\cite{vergassolaetal} A particularly
interesting initial condition which has been much studied (for example
because it arises in cosmology \cite{sheetal,sinai,vergassolaetal}) is
to take for $u(a,0)$ the bilateral Brownian motion curve passing
through the origin (see Fig.~\ref{f:brown}).  For $a\ge 0$ it is
defined as a Gaussian random function with zero mean and with
correlation function $\langle u(a,0)u(a',0)\rangle = \inf (a,\, a')$.
For $a<0$ it is defined as $\tilde u(-a,0)$ where $\tilde u(a,0)$ is
another realization of $u(a,0)$ independent of $u(a,0)$. With such
choice of a self-similar initial condition it is easily shown that the
statistical properties of the solutions at any two given non-vanishing
times are related by a simple change of scale.\cite{sheetal} Thus no
generality is lost by assuming $t=1$, as we shall do here.

\begin{figure}[htbp]
\centerline{\psfig{file=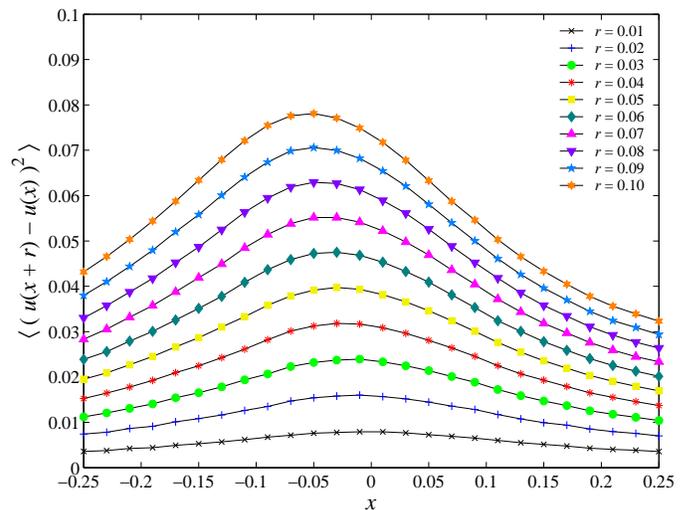,width=0.5\textwidth}}
\caption{The second-order structure function  for various separations $r$,
plotted vs the absolute position $x$. Notice the strong $x$-dependence, 
evidence that incremental homogeneity does not hold. At large $x$ 
the graphs become flat (not shown).}
\label{f:struct}
\end{figure}

In practice $u(a,0)$ is generated as a bilateral random walk (with
steps $\pm\sqrt{\Delta a}$ ) on a regular grid with mesh $\Delta a =
10^{-3}$, covering the Lagrangian interval $[-L_{\rm lag},\, +L_{\rm
lag}]$ where $L_{\rm lag} = 10^3$. The solution is calculated in a
much smaller Eulerian interval $[-L_{\rm eul},\, +L_{\rm eul}]$ with
$L_{\rm eul} = L_{\rm lag}/100 =10$ to ensure that the probability to
have a Lagrangian antecedent outside the interval $[-L_{\rm lag},\,
+L_{\rm lag}]$ is negligibly small.  It is important not to take
initial conditions just from a Fourier series since this would make
the initial condition periodic, thereby loosing incremental
homogeneity in favor of homogeneity, which is dynamically
preserved. Periodicity, assumed in previous numerical studies,
\cite{vergassolaetal} thus cannot  reveal what we see here.  
Fig.~\ref{f:struct} shows the second-order structure function $\langle
\left(u(x+r)-u(x)\right)^2\rangle$ obtained after averaging over
$10^5$ realizations of the initial conditions. It is here plotted
against the absolute position $x$ at fixed separation $r$. A strong
dependence on $x$ reveals that incremental homogeneity does not hold,
except at very large $x$ where the structure function becomes
translation-invariant.  The validity of the numerical results was
tested by increasing by one order of magnitude the length of the
Lagrangian interval and decreasing the number of realizations so that
the computation remains manageable. Exactly the same picture as before
is obtained, but of course with noisier statistics.

A proof of the loss of incremental homogeneity in the temporal
evolution from an initially incrementally homogeneous velocity can be
given for the case of smooth initial data.  The general idea is that
as long as there are no shocks the solution can be Taylor expanded in
time. The second-order structure function can then be calculated
perturbatively and shown not to be translation-invariant. Actually for
Gaussian initial data some of the realizations will develop shocks
after very short times but their contribution to the structure
function can be bounded by terms which are exponentially small in
$1/t^2$. Because of its perturbative nature it is likely that the
proof can be extended to the case of the 3-D Euler or Navier--Stokes
equations. The proof is quite long and technical; its details do not
belong here.

Since incrementally homogeneous random functions may be obtained by
infrared limits of homogeneous random functions for which homogeneity
persists under Burgers or Navier--Stokes dynamics, it is of interest
to understand how homogeneity and incremental homogeneity are lost
when the limit is taken.  For concreteness we shall consider the
Burgers equation with an initial velocity which is the Gaussian
Ornstein--Uhlenbeck process (OUP).  The argument given below is
actually very general and can be applied \emph{mutatis mutandis} to
the Navier--Stokes equation. The OUP $u(x,0;k_0)$ has the spectrum
$1/(k_0^2+k^2)$. Hence its correlation length $L\sim 1/k_0$ and its
variance $\langle u^2(x,0;k_0) \rangle \sim 1/k_0^2$ go to infinity as
$k_0\to 0$. The Brownian motion process must be equal to zero at the
origin. It is obtained from the OUP by first subtracting the
\textit{random} velocity $u(0,0;k_0)$ and then letting $k_0 \to 0$. We
are here emphasizing ``random'' because there is a similar stress put
by Kolmogorov on the first page of his first 1941 paper.  It is clear
that the effect of the initial subtraction of a uniform velocity will
be a (random) Galilean transformation. Hence, denoting by $u(x,t;k_0)$
the solution of the Burgers equation with the initial condition
$u(x,0;k_0)$, the solution of the problem with $u(0,0;k_0)$ subtracted
is
\begin{equation}
v(x,t;k_0) \equiv u\left(x+tu(0,0;k_0),t;k_0\right)-u(0,0;k_0).
\label{galil}
\end{equation}
As $k_0\to 0$ the function $v(x,t;k_0)$ tends to the solution with
Brownian initial condition which was determined numerically above.
Inspecting the r.h.s.\ of (\ref{galil}) we note that if we had only
the subtraction of $u(0,0;k_0)$ the solution and its limit would
remain incrementally homogeneous. However, we also have a dependence
on $u(0,0;k_0)$ inside the spatial argument. Under the assumption of
local homogeneity, the r.h.s. of (\ref{galil}) would be independent of
the velocity $u(0,0;k_0)$ at the reference point (here the origin) and
thus $v$ would be incrementally homogeneous. It is this violation of
local homogeneity (after subtraction of $u(0,0;k_0)$) which spoils
both homogeneity and incremental homogeneity. Local and incremental
homogeneity can only hold in an asymptotic sense at points
sufficiently remote in space-time from the reference point to be only
weakly influenced by its velocity.

At the very beginning of his third 1941 turbulence paper on the
four-fifths law Kolmogorov\cite{k41c} assumed local
homogeneity.\footnote{Actually, he assumed local isotropy, that is
local homogeneity plus isotropy.}  How is the four-fifths law affected
by what we just discussed? Monin and Yaglom \cite{my2} realized that
there may be problems in deriving the four-fifths law (and its
anisotropic generalization) using local homogeneity rather than full
homogeneity. Hence they proposed first (on p.~395) to consider the
locally homogeneous/isotropic turbulence as being embedded in a
large-scale homogeneous turbulence. Then (on pp.~401--403) they
derived the same results for locally homogeneous/isotropic turbulence
using a quasi-Lagrangian method with a frame attached to a particular
fluid particle.  Actually, as we have seen, neither incremental nor
local homogeneity can hold for turbulence which is initially not
strictly homogeneous, except in an asymptotic sense which would not
help in deriving a four-fifths law.  Even in a putative incremental
homogeneous turbulence, attempts to derive a four-fifths law lead to
an extra non-vanishing term as shown in Ref.~\onlinecite{lindborg}
(second term in Eq.~(39)) and in Ref.~\onlinecite{hill02arxiv} (second
term in Eq.~(21)). This term is associated to the advection of
products of two velocity differences by the half sum.  Of course, the
extra term vanishes if homogeneity (or local homogeneity) were to
hold.  We cannot rule out that some conditional form of homogeneity
holds which allows a dependence on the reference velocity but is
consistent with the four-fifths law.

When the turbulence is initially not homogeneous we cannot write a
standard four-fifths law but there is still a relation between the
local fluctuating energy dissipation $\varepsilon({\bm x})$ and the
velocity increments, due to Duchon and Robert.\cite{dr00} In it
simplest form it reads
\begin{equation}
  \varepsilon (\bm x)\! =\! \lim_{r\to 0} \frac{-3}{16\pi
  r^2}\!\!\int\!\! d\Omega\, |\bm u(\bm x + \bm r) - \bm u(\bm x) |^2
  (\bm u(\bm x + \bm r) - \bm u (\bm x))\cdot\bm r, \label{dr}
\end{equation}
where $d\Omega$ is the element of solid angle spanned by the unit
vector ${\bm r}/r$.  This relation, which holds only after the limit
$\nu\to 0$ has been taken, is between non-averaged quantities. In
unpublished notes by Onsager (cited by Eyink and Sreenivasan
\cite{es04}) the same relation appears in averaged
form.\footnote{Onsager seems to have been aware that averaging is not
needed since he crossed out some of the averages in his derivation.}
The Duchon--Robert formula is important because it is an exact
consequence of the dynamical equations and also because it can be
viewed as an infinitesimal version of Kolmogorov's 1962 Refined
Similarity Hypothesis.\cite{k62}

We have seen that Kolomogorov's assumption of independence of
statistical properties of increments on the velocity at the reference
point holds at best in an asymptotic sense, far from the reference
point. More generally, one can ask if the small-scale properties of
turbulent flow near a given point depend much on the instantaneous
velocity $u_0$ at this point.  There is good evidence that it
does. For example, there are recent experimental\cite{cmb04} and then
numerical\cite{bbcdlt04} data showing a strong dependence of the
variance of the fluid particle acceleration, conditioned on $u_0$,
which is found to vary roughly as $u_0^{9/2}$. Indeed as shown in
Ref.~\onlinecite{bbclt04}, large velocity and (extremely) large
acceleration events occur in a correlated way near vortex filaments.

There is nothing surprising to find dependence on $u_0$ in homogeneous
and isotropic turbulence: if in such a flow we observe a high
fluctuation of the local velocity we also expect a high fluctuation in
the local Reynolds number, a sharp decrease in the dissipation scale,
etc.  Some of the dependence on $u_0$ can be captured by adapting an
old argument of Landau (see Ref.\ \onlinecite{ufbook} Sec.\ 6.4.2). In
its original form, Landau's argument assumed that the \emph{mean
dissipation} $\varepsilon$ which appears in the Kolmogorov 1941
expression for the structure functions, can actually be taken as a
quantity presenting large-scale fluctuations.  As pointed out in
Ref.~\onlinecite{frischshe91} arguments \`a la Landau have the
potential of interlinking large-scale and small-scale phenomena. For
this, one just uses the Kolmogorov 1941 predictions in terms of the
large scale velocity $u_0$, the integral scale $L$ and the viscosity
(when needed) and then one allows fluctuations in $u_0$.\footnote{A
somewhat similar procedure was followed by Obukhov when he
reformulated the Kolmogorov 1941 theory to allow fluctuations of
$\varepsilon$ on various scales (A.M.~Obukhov, ``Some specific
features of atmospheric turbulence,'' J.~Fluid Mech.\ \textbf{13}, 77
(1962)).} For example, as pointed out in Ref.~\onlinecite{am04}, from
the Heisenberg--Yaglom theory which predicts that the typical
acceleration of fluid particles is proportional to $u_0^{9/4}$, one
can then infer that its conditional variance goes as $u_0^{9/2}$. If
$u_0$ is assumed to have a Gaussian distribution, the PDF of the
acceleration will involve a stretched exponential.\cite{bbcdlt04} Such
non-Gaussianity should not be mistaken for intermittency. It is just a
consequence of a slight extension of the Kolmogorov theory, which
might be called Kolmogorov--Landau theory.

A delicate issue concerns the effect of $u_0$ on small-scale velocity
increments and in particular on the longitudinal and transverse
second-order structure functions. This question is now being revisited
by various colleagues.

We are grateful to L.~Biferale, E.~Bodenschatz, G.~Eyink, Y.~Gagne,
E.~Lindborg, D.~Mitra, J.-F.~Pinton and K.~Sreenivasan for useful
remarks.  This research was supported by the Indo-French Centre for
the Promotion of Advanced Research (Project 2404-2), by the European
Union under contract HPRN-CT-2002-00300 and by the Swedish Research
Council under contract 2003-4614.


\begin{thebibliography}{99}
\bibitem{kelvin}
Kelvin, Lord (Sir W. Thomson), ``On the propagation of laminar motion
through a turbulently moving inviscid liquid,'' Phil.\ Mag.\
\textbf{24}, 342 (1887).
\bibitem{shiryaev}
A.N.~Shiryaev, ``Kolmogorov and the turbulence'', available at
www.maphysto.dk/publications/MPS-misc/1999/12.pdf .
\bibitem{k40a}
A.N.~Kolmogorov, ``Curves in a Hilbert space that are invariant under
the one-parameter group of motions,'' Dokl. Akad. Nauk SSSR
\textbf{26}, 6 (1940).
\bibitem{k40b}
A.N.~Kolmogorov, ``Wiener's spiral and some interesting curves in
Hilbert space,'' Dokl.\ Akad.\ Nauk SSSR, \textbf{26}, 115 (1940).
\bibitem{k41a}
A.N.~Kolmogorov, ``The local structure of turbulence in incompressible
viscous fluid for very large Reynolds number,'' Dokl.\ Akad.\ Nauk
SSSR \textbf{30}, 9 (1940).
\bibitem{hill97}
R.J.~Hill, ``Applicability of Kolmogorov's and Monin's equations of
turbulence,'' J.\ Fluid Mech.\ \textbf{353}, 67 (1997).
\bibitem{hill02arxiv}
R.J.~Hill, ``The approach of turbulence to the locally homogeneous
asymptote as studied using exact structure-function equations'',
arxiv:physics/0206034 (2002).
\bibitem{my2}eds. 
A.S.~Monin and A.M.~Yaglom, \textit{Statistical Fluid Mechanics},
vol.~2, ed. J.\ Lumley. MIT Press, Cambridge, MA (1975).
\bibitem{k41c}
A.N.~Kolmogorov, ``Dissipation of energy in locally isotropic
turbulence,'' Dokl.\ Akad.\ Nauk SSSR \textbf{32}, 16 (1941).
\bibitem{burgulence}
U.~Frisch and J.~Bec, ``Burgulence'', in \textit{New Trends in
Turbulence}, A.~Yaglom, F.~David and M.~Lesieur eds., Les Houches
Session LXXIV (Springer EDP-Sciences, 2001) p.~341.
\bibitem{vergassolaetal}
M.~Vergassola, B.~Dubrulle, U.~Frisch and A.~Noullez, ``Burgers'
equation, Devil's staircases and the mass distribution for large-scale
structures, ''Astron.\ Astrophys.\ \textbf{289}, 325 (1994).
\bibitem{sheetal}
Z.S.~She, E.~Aurell and U.~Frisch, ``The inviscid Burgers equation
with initial conditions of Brownian type,'' Comm.\ Math.\ Phys.\
\textbf{148}, 623 (1992).
\bibitem{sinai}
Ya.~Sinai, ``Statistics of shocks in solutions of inviscid Burgers
equation,'' Comm.\ Math.\ Phys.\ \textbf{148}, 601 (1992).
\bibitem{lindborg}
E.~Lindborg, ``Horizontal velocity structure functions in the upper
troposphere and lower stratosphere 2. Theoretical considerations,''
J.\ Geophys.\ Res.\ \textbf{106}, 10,233 (2001).
\bibitem{dr00}
J.~Duchon and R.~Robert, ``Inertial energy dissipation for weak
solutions of incompressible Euler and Navier--Stokes equations,''
Nonlinearity \textbf{13}, 249 (2000).
\bibitem{es04}
G.~Eyink and K.~Sreenivasan, ``Onsager and the theory of hydrodynamic
turbulence,'' Rev.\ Mod.\ Phys.\, in press (2005).
\bibitem{k62}
A.N.~Kolmogorov, ``A refinement of previous hypotheses concerning the
local structure of turbulence in a viscous incompressible fluid at
high Reynolds number,'' J.\ Fluid Mech.\ \textbf{13}, 82 (1962).
\bibitem{cmb04}
A.M.~Crawford, N.~Mordant and E.~Bodenschatz, ``Joint statistics of
the Lagrangian acceleration and velocity in fully developed
turbulence,'' Phys.\ Rev.\ Lett.\ \textbf{94}, 024501 (2005).
\bibitem{bbcdlt04}
L.~Biferale, G.~Boffetta, A.~Celani, B.J.~Devenish, A.~Lanotte and
F.~Toschi, ``Multifractal statistics of Lagrangian velocity and
acceleration in turbulence,'' Phys.\ Rev.\ Lett.\ \textbf{93}, 064502
(2004).
\bibitem{bbclt04}
L.~Biferale, G.~Boffetta, A.~Celani, A.~Lanotte and F.~Toschi,
``Particle trapping in three-dimensional fully developed turbulence,''
Phys.\ Fluids \textbf{17}, 021701 (2005).
\bibitem{ufbook}
U.~Frisch, \textit{Turbulence. The Legacy of A.N.~Kolmogorov},
Cambridge University Press, Cambridge (1995).
\bibitem{frischshe91}
U.~Frisch and Z.S.~She, ``On the probability density function of
velocity gradients in fully developed turbulence,'' Fluid Dyn.\ Res.\
\textbf{8}, 139 (1991).
\bibitem{am04}
A.K.~Aringazin and M.I.~Mazhitov, ``Stochastic models of Lagrangian
acceleration of fluid particle in developed turbulence,'' Int.\ J.\
Mod.\ Phys.\ B \textbf{18}, 3095 (2004).
\end{thebibliography}
\end{document}